%% file: vertex2015RD50.tex
\documentclass{PoS}
\usepackage{lineno}
\modulolinenumbers[5]

\title{Simulation of radiation-induced defects}

\ShortTitle{Simulation of radiation-induced defects}

\author{\speaker{Timo Peltola}%
	 \thanks{On behalf of the RD50 Collaboration.}\\
        Helsinki Institute of Physics\\
        E-mail: \email{timo.peltola@helsinki.fi}}

\abstract{Mainly due to their outstanding performance the position sensitive silicon detectors are widely used in the tracking systems of High Energy Physics experiments such as the ALICE, ATLAS, CMS and LHCb at LHC, the world's largest particle physics accelerator at CERN, Geneva. The foreseen upgrade of the LHC to its high luminosity (HL) phase (HL-LHC scheduled for 2023), will enable the use of maximal physics potential of the facility. After 10 years of operation the expected fluence 
will expose the tracking systems at HL-LHC to a radiation environment that is beyond the capacity of the present system design. Thus, for the required upgrade of the all-silicon central trackers extensive measurements and simulation studies for silicon sensors of different designs and materials with sufficient radiation tolerance have been initiated within the RD50 Collaboration.

Supplementing measurements, simulations are in vital role for e.g. device structure optimization or predicting the electric fields and trapping in the silicon sensors. 
The main objective of the device simulations in the RD50 Collaboration is to develop an approach to model and predict the performance of the irradiated silicon detectors 
using professional software.
The first successfully developed quantitative models for radiation damage, based on two effective midgap levels, are able to reproduce the experimentally observed detector characteristics like leakage current, full depletion voltage and charge collection efficiency (CCE). 
Recent implementations of additional traps at the SiO$_2$/Si interface or close to it have expanded the scope of the experimentally agreeing simulations to such surface properties as the interstrip resistance and capacitance, and the position dependency of CCE for strip sensors irradiated up to $\sim$$1.5\times10^{15}$ n$_{\textrm{\small eq}}\textrm{cm}^{-2}$.

Insights to the development processes of the defect models as well as a review of the recent results from 
TCAD simulations of several detector technologies and silicon
materials at radiation levels expected for HL-LHC will be presented.}

\FullConference{24th International Workshop on Vertex Detectors\\
                 1-5 June 2015\\
                 Santa Fe, New Mexico, USA}

\begin{document}

\section{Introduction}
\label{Introduction}
\input{Introduction.tex}

\section{Radiation induced defects}
\label{RadiationInducedDefects}
\input{Defects.tex}

\section{Simulated defects}
\label{SimulatedDefects}
The TCAD simulations in the following sections applied both n- and p-type bulk material with $\langle 1,0,0\rangle$ crystal orientation. The modeled proton and neutron irradiations were always considered in 1 MeV neutron equivalent fluences (n$_{\textrm{\small eq}}$cm$^{-2}$). 

\subsection{Implementation of defects - From PTI model to TCAD models}
\label{Implementation}
\input{Implementation.tex}
\subsection{Bulk damage in TCAD}
\label{BulkDamage}
\input{BulkD.tex}
\subsection{TCAD models with bulk \& surface damage}
\label{SurfaceDamage}
\input{SurfaceD.tex}

\section{Summary \& outlook}
\label{Summary}
\input{Summary.tex}
\acknowledgments{Author would like to thank the colleagues of the RD50 Collaboration for the material and support.}

\end{document}

%% file: Introduction.tex
Tracking systems of High Energy Physics experiments largely employ position sensitive silicon sensors due to their outstanding performance and cost effectiveness. 
They are currently installed in the vertex and tracking detectors of the ALICE, ATLAS, CMS and LHCb experiments at LHC, the world's
largest particle physics accelerator at CERN. An upgrade of LHC accelerator is already planned for 2023, namely the high luminosity phase of
the LHC (HL-LHC). Approximately a 10-fold increase of the luminosity will enable the use of maximal physics potential of the machine.
After 10 years of operation, the integrated luminosity of 3000 fb$^{-1}$ \cite{bib1,bib2,bib3}
will expose the tracking systems at HL-LHC to a radiation environment that is beyond the capability of the present system design. At ATLAS and CMS fluences of more than $1\times10^{16}$ $\textrm{n}_{\textrm{\small eq}}$cm$^{-2}$ (1 MeV neutron equivalent) are expected for the pixel detectors at the innermost layers and above $10^{15}$ $\textrm{n}_{\textrm{\small eq}}$cm$^{-2}$ for the strip sensors $\sim$20 cm from vertex \cite{bib4}. 
In addition to the extremely
high radiation levels increase in the track density will be among the most demanding challenges. 

Therefore a dedicated R$\&$D program is needed to 
improve the present detector technologies, or develop novel ones,
for both the innermost tracking layers and 
most of the outer tracker components with detectors that can endure higher radiation levels and higher occupancies.
The RD50 collaboration "Development of Radiation Hard Semiconductor Devices for Very High Luminosity
Colliders" \cite{bib5,bib6} was formed in 2002 with the objective to develop
semiconductor sensors that meet the HL-LHC requirements mentioned above. 

Alongside with measurements, the simulations are an integral part of the detector development process. Their role is especially vital in areas like device structure optimization or predicting the electric fields and trapping in the silicon sensors. 
When the numerical simulations are capable to verify experimental results they will also gain predictive power, resulting in reduced time and cost budget in detector design and testing. 
The main focus of the device simulations in the RD50 Collaboration is to develop an approach to model and predict the performance of the irradiated silicon detectors (diode, strip, pixel, columnar 3D) using professional software, namely finite-element Technology Computer-Aided Design (TCAD) software frameworks Synopsys Sentaurus\footnote{http://www.synopsys.com} and Silvaco Atlas\footnote{http://www.silvaco.com}. Among the multitude of simulation options the TCAD packages allow the incorporation of realistic, segmented sensors in 2D or 3D, readout circuit, transient simulations with lasers or Minimum Ionizing Particles (MIPs), etc.

In the following an insight to the development path from observed defects to the effective defect model and from the initial 1-dimensional model for custom-made software to the quantitative TCAD models is given as well as selected recent 
simulation results and their comparison with measurements of several detector technologies and silicon materials at radiation levels expected for the HL-LHC.

%% file: Defects.tex
Radiation levels above $\sim$$10^{13}$ $\textrm{n}_{\textrm{\small eq}}$cm$^{-2}$ cause damage to the silicon crystal structure. Fluences above $1\times10^{14}$ $\textrm{n}_{\textrm{\small eq}}$cm$^{-2}$ lead to significant degradation of the detector performance. In the intense radiation fields of the LHC, defects are introduced both in the silicon substrate (bulk damage) and in the SiO$_2$ passivation layer, that affect the sensor performance through the interface with the silicon bulk (surface damage). Bulk damage degrades detector operation by introducing deep acceptor and donor type trap levels \cite{bib7}. The main macroscopic effects of bulk damage on high-resistivity silicon detectors irradiated by hadrons are the change of the effective doping concentration $N_{\textrm{\small eff}}$, the increase in the leakage
current proportional to the fluence and the degradation of Charge Collection Efficiency (CCE) \cite{bib8}. 

Surface damage consists of a positively charged layer accumulated inside the oxide and of interface traps 
created close to the interface with silicon bulk \cite{bib9,bib10}. 
High oxide charge densities $Q_{\textrm{\small f}}$ are detrimental to the detector performance since the electron layer generated under the SiO$_2$/Si interface can cause very high electric fields $E$($x$) near the p$^+$ strips in p-on-n sensors and loss of position resolution in n-on-p sensors by providing a conduction channel between the strips. High $E$($x$) can induce detector breakdown or avalanches that can result in non-Gaussian noise events. Also the increase of interstrip capacitance $C_{\textrm{\small int}}$ with the accumulating interface charges will contribute to higher strip noise.

Both point defects and clusters are responsible for the various damage effects in the detector
bulk, depending on their concentration, energy level and the respective electron and
hole capture cross-section. Low-energy recoils above specific particle threshold energies will usually create fixed point
defects. At the recoil energies above $\sim$5 keV a dense agglomeration of defects is formed at the end of the primary Primary Knock-on Atom (PKA) track. 
These disordered regions are referred to as defect clusters \cite{bib11,bib12}. Measurements with methods like Thermally Stimulated 
Current technique (TSC), Deep Level Transient Spectroscopy (DLTS), Transient Current Technique (TCT), Capacitance-Voltage (C-V) and Current-Voltage (I-V) have revealed a multitude of defects (11 different energy levels listed in \cite{bib13}) after irradiations with hadrons or higher energy leptons. The microscopic defects have been observed to influence the macroscopic properties of a silicon sensor by 
charged defects contributing 
to the effective doping concentration \cite{bib14,bib15,bib16,bib17,bib18,bib19} and deeper levels also to trapping and generation/recombination of the charge carriers (leakage current) \cite{bib19,bib20,bib21,bib22}. 
Such quantity of defect levels set up a vast parameter space that is 
neither practical nor purposeful to model and tune. Thus, a minimized set of defects constituting an effective defect model is the most meaningful approach for the simulations of irradiated silicon detectors.

%% file: Implementation.tex
The information of the defects discussed in the previous section is used as a starting point to the device simulations. 
The simulation of radiation damage in the silicon bulk is based on the effective mid gap levels (a deep acceptor and a deep donor level with activation energies $E_{\textrm{\small a}}$ = $E_{\textrm{\small c}}$ - (0.525 $\pm$ 0.025) eV and $E_{\textrm{\small v}}$ + 0.48 eV, respectively). The model, presented in table~\ref{tab:1}, was first proposed in 2001 by Eremin, Verbitskaya \& Li, and entitled later as the "PTI model" (also "EVL model") \cite{bib23,bib24,bib25}.  
The main idea of the model is that the two peaks in the electric field profile $E$($z$) 
of the both proton and neutron irradiated detectors can be explained via the interaction of the carriers from the bulk generated current with the electron traps and simultaneously with the hole traps. 
The physical meaning behind all later models is the same combination of the hole and electron traps. 

\begin{figure}[tbp] 
\centering
\includegraphics[width=.5\textwidth]{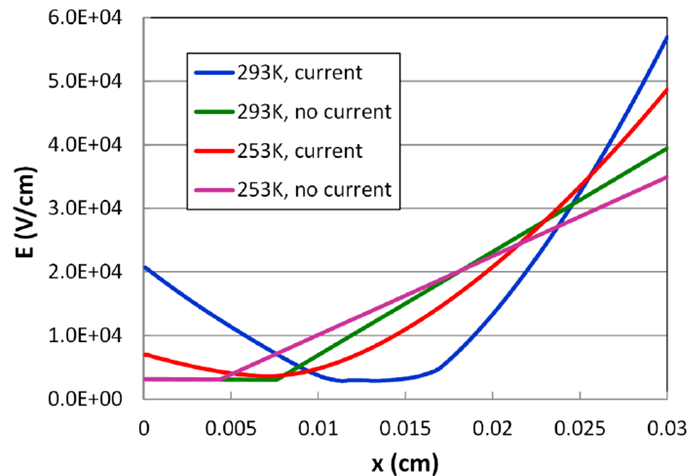}
\caption{The dependence of the electric field profile in the Si bulk on the current generating level of the PTI model. The 300 $\mu\textrm{m}$ thick 1-dim. diode structure was set to irradiation fluence of $1\times10^{15}$ $\textrm{n}_{\textrm{\small eq}}$cm$^{-2}$ at $V$ = 500 V. The double peak formation is observed only when the two midgap levels are supplemented by the current generating level \cite{bib26}.}
\label{fig:8}
\end{figure}

Simulation by a custom-made software of a 1-dim. diode structure presented in figure~\ref{fig:8} shows the double peak formation and its dependence on the current generating level. 
%
%
Since the PTI model was not designed for the TCAD simulation packages and it is basically a three trap model with the two deep traps used to create additional space charges and the third middle level to generate leakage current only, its adaptation to TCAD is not straightforward. 
To avoid the artificial trap level which generates exclusively leakage current and to account for the experimentally measured leakage current, modifications for the implementation of the defect model to the Synopsys Sentaurus and Silvaco Atlas packages are required.
\begin{table}[tbp]
\smallskip
\centering
\begin{tabular}{|c|c|c|c|}
    \hline
    {\bf Defect type} & {\bf Level} \textrm{[eV]} & {\bf $\sigma_{\textrm{\small e,h}}$} \textrm{[cm$^{2}$]} & 
{\bf $\eta$} \textrm{[cm$^{-1}$]}\\
    \hline
    Deep acceptor &  $E_{\textrm{\small c}}-0.525$ & $1\times10^{-15}$ & 1\\
    Deep donor &  $E_{\textrm{\small v}}+0.48$ & $1\times10^{-15}$ & 1\\
    Current generating level &  $E_{\textrm{\small c}}-0.65$ & $1\times10^{-13}$ & 1\\
    \hline
\end{tabular}
\caption{The initial parameters of the PTI model \cite{bib23}. $E_{\textrm{\small c,v}}$ are the conduction band and valence band energies, $\sigma$$_{\textrm{\small e,h}}$ are the electron and hole trapping cross sections and $\eta$ is the introduction rate. The defect concentration is given by the product of fluence and $\eta$.}
\label{tab:1}
\end{table}

%% file: BulkD.tex
In the TCAD simulation packages the leakage current can be generated either via tuning the charge carrier lifetimes or by introducing defect levels for the non-irradiated and irradiated sensors, respectively. 
In addition to the leakage current, the lifetime tuning 
affects also the charge collection via modified trapping in the bulk 
while the defect levels simultaneously affect the space charge and trapping.
Thus, to build a TCAD defect model on the base of the PTI model, with the same two deep levels as those used for the double peak $E$($z$) explanation,  
all parameters (except the energy level in the silicon bandgap) have to be determined in agreement with experimental data to correctly describe leakage current $I_{\textrm{\small leak}}$, full depletion voltage $V_{\textrm{\small fd}}$, transient pulses and CCE after irradiation. 
%

\begin{figure}[tbp] 
\centering
\includegraphics[trim=0.0cm 0.0cm 0.0cm 0.8cm, clip=true, width=.48\textwidth]{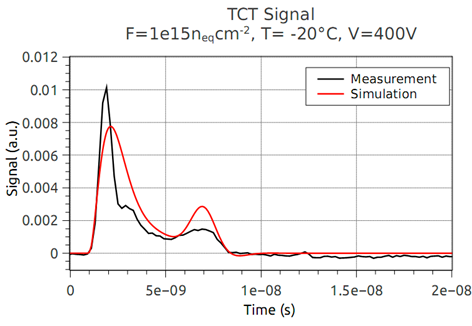}
\includegraphics[width=.48\textwidth]{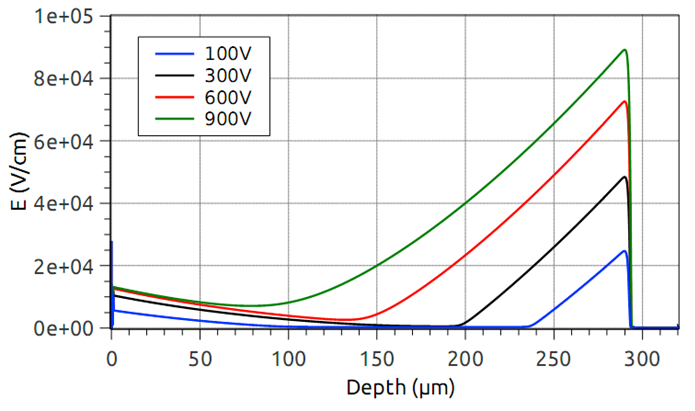}
\caption{(left) Measured and Sentaurus proton model simulated transient signals in a 300 $\mu\textrm{m}$ thick pad detector irradiated by the fluence of $1\times10^{15}\textrm{n}_{\textrm{\small eq}}$cm$^{-2}$ at 
$V$ = 400 V. (right) Corresponding electric field profiles in the detector bulk for varying bias voltages \cite{bib27}.}
\label{fig:9}
\end{figure}

The tuning procedure for the Synopsys Sentaurus defect models, producing bulk properties quantitatively matching the measurements, is described in detail in references \cite{bib27,bib28}. To give a general idea of the process flow, the tuning is realized by first applying parametrization of the generated current at fixed $T=253~\textrm{K}$. 
Next, the C-V curves are tuned to match experimental results 
by iterating the ratio of donor concentration to acceptor concentration. 
The significant difference in measured $V_{\textrm{\small fd}}$($\Phi_{\textrm{\small eq}}$) for proton and neutron irradiated detectors requires exclusive defect models for each radiation type. Finally, by tuning the acceptor and donor capture cross sections, the transient current curves are adjusted to match the measured signals, shown on the left plot of figure~\ref{fig:9}. 
This is necessary for the simulated CCE to reflect trapping in a real irradiated detector. The models, namely the proton model and the neutron model, are valid for fluences from $10^{14}$ to $\sim$$1.5\times10^{15}$ n$_{\textrm{\small eq}}$cm$^{-2}$ with parameters presented in tables~\ref{tab:2} and~\ref{tab3}. 

Tuning of the Silvaco Atlas defect model for proton irradiation 
\cite{bib29}, includes also the modification of the deep acceptor energy level value within the PTI model error bars, presented in table~\ref{tab:4}, and it reproduces the expected $I_{\textrm{\small leak}}$, $V_{\textrm{\small fd}}$ and $E$($z$) up to fluences $\sim$10$^{15}$ n$_{\textrm{\small eq}}$cm$^{-2}$. Earlier defect models involving solely bulk damage, include a two-level model with tuned parameters from the PTI model \cite{bib30}, a three-level model with different energy levels to PTI model \cite{bib31}, and its capture cross-section tuned version \cite{bib32}. 

\begin{figure}[tbp] 
\centering
\includegraphics[width=.44\textwidth]{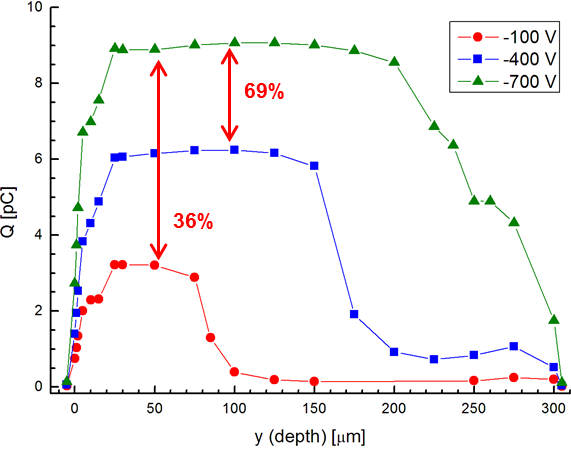}
\includegraphics[width=.5\textwidth]{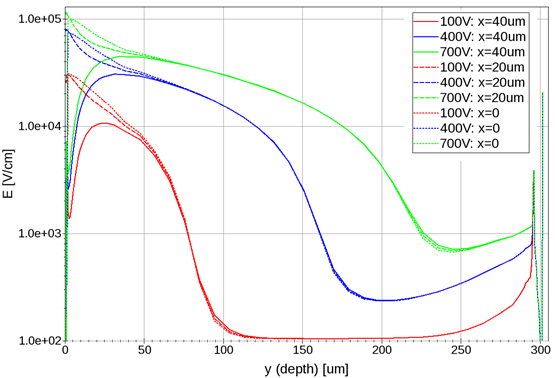}
\caption{(left) Edge-TCT simulation using the Sentaurus neutron model with matching detector parameters to a measurement by G. Kramberger et al. \cite{bib33} of a neutron irradiated 300 $\mu\textrm{m}$ thick n-on-p strip detector at the fluence of $5\times10^{14}$ cm$^{-2}$ 
\cite{bib28}. Corresponding measured ratios of the peak collected charges were $\sim$$33\%$ and $\sim$$71\%$. (right) Corresponding electric field distributions in the Si bulk with cuts made from the center of the strip ($x=0$) to the center of the interstrip gap ($x=40~\mu\textrm{m}$).} 
\label{fig:10}
\end{figure}

The reliability of the simulated electric field distributions, which can not be measured directly, can be confirmed by comparing the measured and simulated transient current curves and collected charges from surface (TCT) and side-plane (edge-TCT) charge injections. The former method is presented in the two plots of figure~\ref{fig:9} where after red laser front surface illumination the resulting carrier drift in the double peak electric field is reflected by a double peak in the transient signal. 
Presented on the left plot of figure~\ref{fig:10} are the collected charges in a neutron irradiated 300 $\mu\textrm{m}$ thick n-on-p strip detector after infrared laser injections from the side-plane of the sensor at various depths, i.e. an edge-TCT simulation. 
%
Both the relative peak values of the collected charges between voltages and the extension of the depletion region (high collected charges), especially for the higher voltages, are in agreement with the measurements in reference \cite{bib33}. When also the double peak formation observed in the measurement is reproduced by the simulation, the electric field distributions on the right plot of figure~\ref{fig:10} can be considered to model the real sensor reliably.

%
Electron trapping time in the Sentaurus simulation can be determined from the time evolution of the integrated electron density, described in detail in \cite{bib27}. Its value for proton model simulation at the fluence of e.g. $1\times10^{14}\textrm{n}_{\textrm{\small eq}}$cm$^{-2}$ was calculated as $\tau_{\textrm{\small e}}=28.2\pm0.6~\textrm{ns}$.  While this is $\sim30\%$ above the experimental value in reference \cite{bib34}, it can be deemed a decent preliminary result when considering the error sources: 10 K higher temperature in the measurement and 
shorter TCT pulse in the simulation. In a shorter pulse, smaller number of charge carriers experience trapping. 
%
%

\begin{figure}[tbp] 
\centering
\includegraphics[width=.46\textwidth]{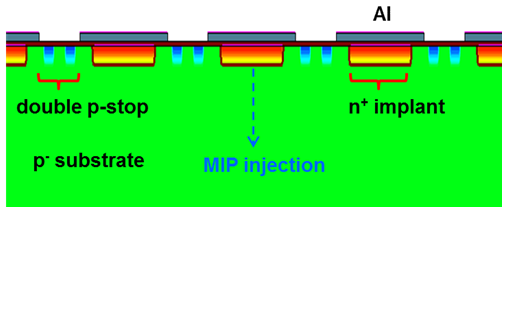}
\includegraphics[width=.5\textwidth]{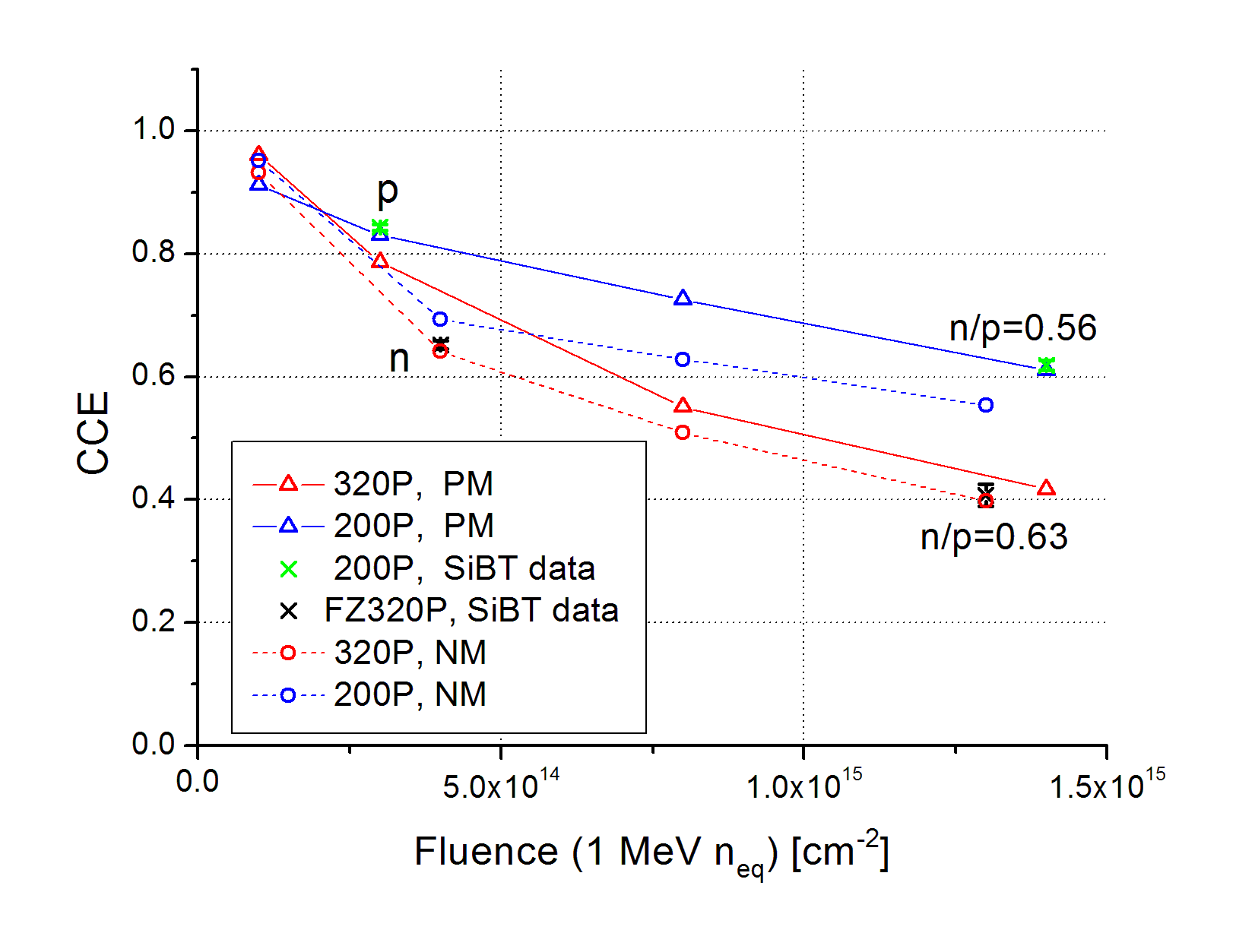}
\caption{(left) Sentaurus generated n-on-p 5-strip sensor front surface with a double p-stop strip isolation structure (not to scale). For the CCE simulations the charge was injected from the middle of the centermost strip. 
(right) Measured and simulated CCE($\Phi_{\textrm{\small eq}}$) for the n-on-p strip sensors with a strip pitch of 120 $\mu\textrm{m}$ \cite{bib35}. Types of irradiation are marked in the plot as p = proton, n = neutron and n/p = mixed fluence with the ratios of the particles indicated. PM = proton model, NM = neutron model and e.g. 200P is a p-type sensor with a 200 $\mu\textrm{m}$ active region thickness. The experimental data was measured with the SiBT set-up \cite{bib37}.}
\label{fig:12}
\end{figure}

As shown in the right plot of figure~\ref{fig:12} the simulated CCE of a 300 and 200 $\mu\textrm{m}$ active thickness sensors compares well to the measured data from both neutron and proton irradiated strip detectors \cite{bib35,bib27,bib28}. At this stage the surface damage was modelled in the Sentaurus simulation by placing a fixed charge density $Q_{\textrm{\small f}}$ at the SiO$_2$/Si interface. By using $Q_{\textrm{\small f}}$ as a further tuning parameter (described in detail in reference \cite{bib35}) to find CCE matching with the measurement for similar sensor structures and equal irradiation type, the fixed $Q_{\textrm{\small f}}$ values can then be applied to make a prediction of CCE($\Phi$) for the sensors with different active thicknesses and equal irradiation type. However, the maximal $Q_{\textrm{\small f}}$ values had to be limited to considerably lower than expected in a real sensor for the highest fluences ($1 - 2\times10^{12}$ $\textrm{cm}^{-2}$ \cite{bib10}, \cite{bib36}) to maintain strip isolation in the proton model simulation. 
%
%
\begin{table}[tbp]
\smallskip
\centering
\begin{tabular}{|c|c|c|c|}
    \hline
    {\bf Defect type} & {\bf Level} \textrm{[eV]} & {\bf $\sigma_{\textrm{\small e,h}}$} \textrm{[cm$^{2}$]} & 
{\bf Concentration} \textrm{[cm$^{-3}$]}\\
    \hline
    Deep acceptor &  $E_{\textrm{\small c}}-0.525$ & $1\times10^{-14}$ & $1.189\times\Phi+6.454\times10^{13}$\\
    Deep donor &  $E_{\textrm{\small v}}+0.48$ & $1\times10^{-14}$ & $5.598\times\Phi-3.959\times10^{14}$\\
    \hline
\end{tabular}
\caption{The parameters of the proton model for Synopsys Sentaurus \cite{bib27,bib28}. $E_{\textrm{\small c,v}}$ are the conduction band and valence band energies, $\sigma$$_{\textrm{\small e,h}}$ are the electron and hole trapping cross sections and $\Phi$ is the fluence.} 
\label{tab:2}
\end{table}
\begin{table}[tbp]
\smallskip
\centering
\begin{tabular}{|c|c|c|c|}
    \hline
    {\bf Defect type} & {\bf Level} \textrm{[eV]} & {\bf $\sigma_{\textrm{\small e,h}}$} \textrm{[cm$^{2}$]} & 
{\bf Concentration} \textrm{[cm$^{-3}$]}\\
    \hline
    Deep acceptor &  $E_{\textrm{\small c}}-0.525$ & $1.2\times10^{-14}$ & $1.55\times\Phi$\\
    Deep donor &  $E_{\textrm{\small v}}+0.48$ & $1.2\times10^{-14}$ & $1.395\times\Phi$\\
    \hline
\end{tabular}
\caption{The parameters of the neutron model for Synopsys Sentaurus \cite{bib27,bib28}. Symbols are as in table~\protect\ref{tab:2}.} 
\label{tab3}
\end{table}
\begin{table}[tbp]
\smallskip
\centering
\begin{tabular}{|c|c|c|c|c|}
    \hline
    {\bf Defect type} & {\bf Level} \textrm{[eV]} & {\bf $\sigma_{\textrm{\small e}}$} \textrm{[cm$^{2}$]} & {\bf $\sigma_{\textrm{\small h}}$} \textnormal{[cm$^{2}$]} & 
{\bf Concentration} \textrm{[cm$^{-3}$]}\\
    \hline
    Deep acceptor &  $E_{\textrm{\small c}}-0.51$ & $2\times10^{-14}$ & $2.6\times10^{-14}$ & $4\times\Phi$\\
    Deep donor &  $E_{\textrm{\small v}}+0.48$ & $2\times10^{-14}$ & $2\times10^{-14}$ & $3\times\Phi$\\
    \hline
\end{tabular}
\caption{The parameters of the Delhi University bulk defect model for Silvaco Atlas \cite{bib29}. Symbols are as in table~\protect\ref{tab:2}.} 
\label{tab:4}
\end{table}
%
%

%% file: SurfaceD.tex
Investigation of the TCAD bulk damage models in segmented devices with surface damage modeled by Si/SiO$_2$ interface charge density $Q_{\textrm{\small f}}$ has revealed that the approach is not sufficient to reproduce the observed surface properties of irradiated detectors. As discussed at the end of the section~\ref{BulkDamage}, the strips became shorted at high proton fluences when they were experimentally expected to be isolated. Simultaneously the resulting interstrip capacitances $C_{\textrm{\small int}}$ remained several orders of magnitude above the geometrical values expected from the measurements. Additionally the observed position dependence of CCE in irradiated strip detectors \cite{bib38}, i.e. CCE($x$), was not reproduced also at a lower fluence of $3\times10^{14}$ $\textrm{n}_{\textrm{\small eq}}$cm$^{-2}$. 
%
Thus, these observations represented a further demand for the parameter tuning of the defect models for segmented sensors. 
This has been realized by implementing additional traps $N_{\textrm{\small it}}$ at the SiO$_2$/Si interface (Atlas \cite{bib29}) or extending from it with a wider depth distribution (Sentaurus \cite{bib28,bib35,bib39}). 
Earlier published simulation studies for surface damage only 
have modeled it 
either by interface charge density $Q_{\textrm{\small f}}$ \cite{bib40,bib41} (as in section~\ref{BulkDamage}), by including in the threshold voltage expression the induced flat-band voltage shift \cite{bib42} or by three interface traps with parameters matching the measurements of X-ray irradiated devices \cite{bib9}. 

\begin{figure}[tbp] 
\centering
\includegraphics[width=.42\textwidth]{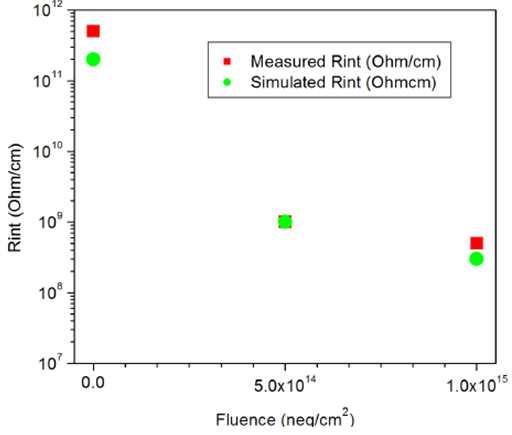}
\includegraphics[width=.48\textwidth]{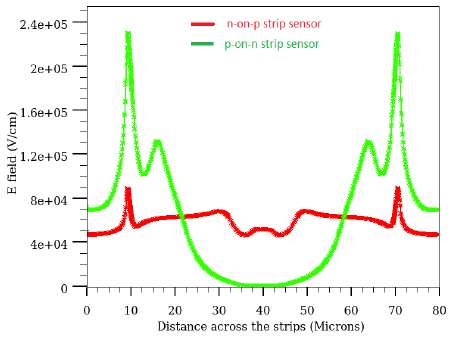}
\caption{Simulation results of Silvaco Atlas defect model with 2 bulk defect levels and interface traps \cite{bib29}. (left) Simulated and measured $R_{\textrm{\small int}}$ as a fuction of fluence for n-on-p strip sensors with double p-stop isolation at $V$ = 600 V. The $R_{\textrm{\small int}}$ simulations for fluences 0, $5\times10^{14}$ and $1\times10^{15}$ n$_{\textrm{\small eq}}$cm$^{-2}$ were carried out using $Q_{\textrm{\small f}}$ (equal to $N_{\textrm{\small it}}$) values of $10^{11}$, $10^{12}$ and $1.5\times10^{12}$ cm$^{-2}$, respectively. (right) Electric field distribution of the n-on-p and p-on-n strip sensors at a cut line 1.4 $\mu\textrm{m}$ below the Si/SiO$_2$ interface at the bias voltage of 500 V and fluence of $1\times10^{15}$ n$_{\textrm{\small eq}}\textrm{cm}^{-2}$.}
\label{fig:13}
\end{figure}

In the Silvaco Atlas model \cite{bib29}, presented in table~\ref{tab:5}, 
%
it is assumed that for a given interface
trap density $N_{\textrm{\small it}}$ 60\% are deep acceptor traps ($E_{\textrm{\small c}}$ - 0.60 eV) and 40\% shallow acceptor traps ($E_{\textrm{\small c}}$ - 0.39 eV). 
Since the measured $Q_{\textrm{\small f}}$ and $N_{\textrm{\small it}}$ values \cite{bib10} are quite similar in magnitude, 
$N_{\textrm{\small it}}$ was set to equal value with $Q_{\textrm{\small f}}$ for the simulations. Complemented by the 2-defect bulk model with experimentally matching properties described in section~\ref{BulkDamage}, the combined bulk \& surface damage model reproduces interstrip resistances $R_{\textrm{\small int}}$ in very close agreement with the measurements, as shown in the left plot of figure~\ref{fig:13}. This enables e.g. a reliable simulation investigation of the electric field distributions between the strips in irradiated n-on-p and p-on-n sensors, presented in the right plot of figure~\ref{fig:13}.
%

When the $N_{\textrm{\small it}}$ approach was applied with the Sentaurus proton model it was found that the expected $C_{\textrm{\small int}}$ values were not reproduced \cite{bib43} at the $Q_{\textrm{\small f}}$ range where CCE($x$) was matching the measurement. By applying a deeper distribution for the surface traps, namely 2 $\mu\textrm{m}$ from the Si/SiO$_2$ interface, it was possible to reach agreement with the measurement for the aforementioned surface properties and maintain strip isolation also at high proton fluences, while leaving the bulk properties of the proton model unaffected \cite{bib35,bib39}. It should be emphasized that the depth distribution of the applied single shallow acceptor level ($E_{\textrm{\small c}}$ - 0.40 eV) is concentration dependent and other values could be used to same effect.  

\begin{figure}[tbp] 
\centering
\includegraphics[width=.44\textwidth]{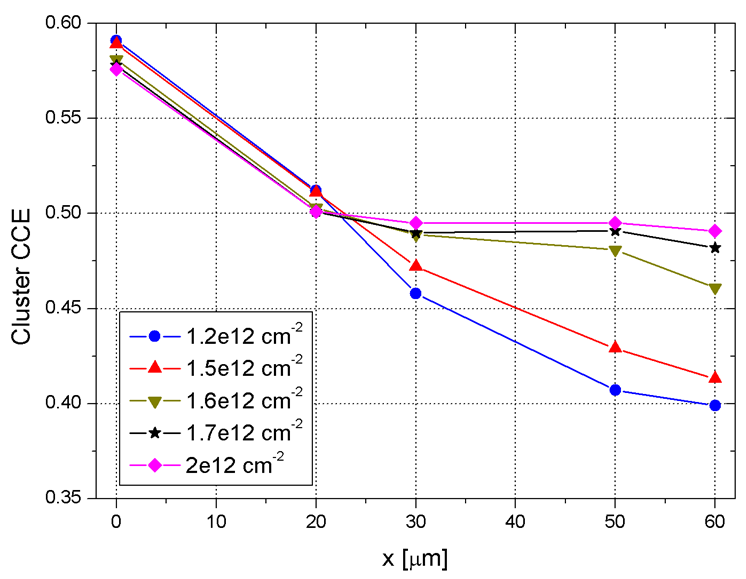}
\includegraphics[width=.45\textwidth]{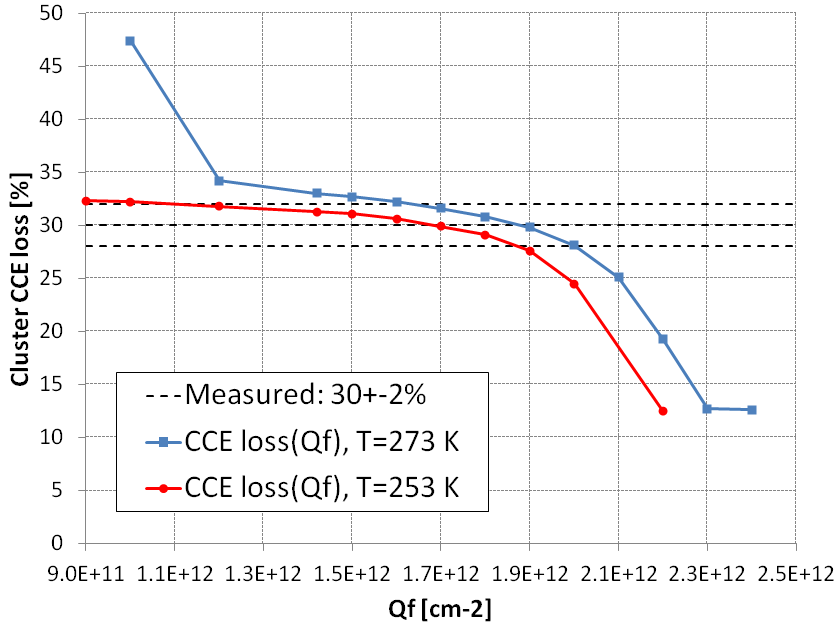}
\caption{Simulation results of a 200 $\mu\textrm{m}$ active thickness n-on-p strip sensor with a 120 $\mu\textrm{m}$ pitch using Synopsys Sentaurus non-uniform 3-level defect model \cite{bib35}. (left) Simulated position dependency of the CCE 
at $V=\textrm{-}1~\textrm{kV}$ and $T=253~\textrm{K}$ after irradiation to $\Phi_{\textrm{\small eq}}=1.5\times10^{15}$ $\textrm{cm}^{-2}$. 
Charge injection positions $x=0$, 60 $\mu\textrm{m}$ correspond to the center of the strip and the center of the interstrip gap, respectively. Varying the values of 
$Q_{\textrm{\small f}}$ displays the evolution of the CCE loss between the strips. 
(right) Measured \cite{bib38} and 
simulated CCE loss 
at the fluence of $\Phi_{\textrm{\small eq}}=1.4\times10^{15}$ $\textrm{cm}^{-2}$ and $V=\textrm{-}1~\textrm{kV}$. Simulation matches the 
measurement 
when $Q_{\textrm{\small f}}=(1.6\pm0.2)\times10^{12}$ $\textrm{cm}^{-2}$. The detectors measured with the SiBT set-up were irradiated by mixed fluences.} 
\label{fig:14}
\end{figure}

While for the CCE simulations the charge injection position was fixed in the middle of the centermost strip, shown on the left side of figure~\ref{fig:12}, for the CCE($x$) simulations the position was varied from the 
midgap between the strips to the center of the strip, presented on the left plot of figure~\ref{fig:14}. This plot also provides information on the strip isolation; when the strips are isolated the cluster CCE decreases towards the midgap but when shorted the cluster CCE becomes position independent. 
The cluster CCE loss between the strips was then tuned to find agreement with the measured value by scanning the interface charge values, illustrated on the right plot of figure~\ref{fig:14}. 
%

Simulated CCE($x$) was found to be dependent on the shallow acceptor concentration in the 3-level defect model and on $Q_{\textrm{\small f}}$ at a given fluence. Thus, by fixing one it is possible to parametrize the other as a function of fluence. 
Due to shortage of exact measured data of $Q_{\textrm{\small f}}$ estimated values were used, against which the shallow acceptor concentration was tuned. Hence, at this stage the approach provides more a method than quantitative information on this regard. 
With the existing measured CCE($x$) data a preliminary parametrization, presented in table~\ref{tab:6}, of the 'non-uniform 3-level model' 
was done for fluences from $3\times10^{14}$ to $1.4\times10^{15}$ n$_{\textrm{\small eq}}$cm$^{-2}$ for the strip pitch of 120 $\mu\textrm{m}$ \cite{bib35}. 
%
%
\begin{table}[tbp]
\smallskip
\centering
\begin{tabular}{|c|c|c|c|}
    \hline
    {\bf Defect type} & {\bf Level} \textrm{[eV]} & {\bf $\sigma_{\textrm{\small e,h}}$} \textrm{[cm$^{2}$]} & 
{\bf Density} \textrm{[cm$^{-2}$]}\\
    \hline
    Deep acceptor &  $E_{\textrm{\small c}}-0.60$ & $1\times10^{-15}$ & $0.6\times Q_{\textrm{\small f}}$\\
    Shallow acceptor &  $E_{\textrm{\small c}}-0.39$ & $1\times10^{-15}$ & $0.4\times Q_{\textrm{\small f}}$\\
    \hline
\end{tabular}
\caption{The parameters of the Delhi University interface trap model for Silvaco Atlas \cite{bib29}. $Q_{\textrm{\small f}}$ is the interface charge density while the rest of the symbols are as in table~\protect\ref{tab:2}.} 
\label{tab:5}
\end{table}
\begin{table}[tbp]
\smallskip
\centering
\begin{tabular}{|c|c|c|c|c|}
    \hline
    {\bf Defect type} & {\bf Level} \textnormal{[eV]} & {\bf $\sigma$$_{\textnormal{\small e}}$} \textnormal{[cm$^{2}$]} & {\bf $\sigma$$_{\textnormal{\small h}}$} \textnormal{[cm$^{2}$]} & {\bf Concentration} \textnormal{[cm$^{-3}$]}\\
    \hline
    Shallow acceptor & $E_{\textnormal{\small c}}-0.40$ & 8$\times$10$^{-15}$ & 2$\times$10$^{-14}$ & 14.417$\times$$\Phi$+3.168$\times$10$^{16}$\\
    \hline
\end{tabular}
\caption{
Parameters of the shallow acceptor level included in the non-uniform 3-level model for Synopsys Sentaurus \cite{bib35}, parametrized for the fluence range $(0.3 - 1.4)\times10^{15}$ n$_{\textrm{\small eq}}\textrm{cm}^{-2}$. Symbols are as in table~\protect\ref{tab:2}.} 
\label{tab:6}
\end{table}
%
%

%% file: Summary.tex
An overview of the RD50 collaboration 
defect simulation activities 
has been given with several examples of the simulation results by the Synopsys Sentaurus and Silvaco Atlas TCAD defect models including comparisons with measurements. 

Data of the measured defects and understanding of their microscopic properties has been essential in the development of the defect models. The multitude of the observed defects in irradiated silicon and the ensuing vast parameter space has led to the approach of minimized set of effective defect levels. Thus, the simulation of the radiation induced damage in silicon bulk is based on the deep acceptor and deep donor levels with the activation energies $E_{\textrm{\small c}}$ - (0.525 $\pm$ 0.025) eV and $E_{\textrm{\small v}}$ + 0.48 eV, respectively. The main concept of the model is that the two peaks in the $E$($z$) profile of both proton and neutron irradiated detectors is explained via the interaction of the charge carriers from bulk generated current simultaneously with electron traps and hole traps.

First succesful quantitative TCAD defect models reproduce experimentally observed leakage current, full depletion voltage, transient signals and CCE at the fluences from $10^{14}$ to $\sim$10$^{15}$ n$_{\textrm{\small eq}}$cm$^{-2}$.
The problematic simulation of the surface properties in heavily irradiated segmented sensors was solved by implementating additional traps at the SiO$_2$/Si interface or close to it expanding the scope of the experimentally agreeing simulations to include 
interstrip resistance and interstrip capacitance. Also the position dependency of CCE for the strip sensors with strip pitch of 120 $\mu\textrm{m}$ was succesfully reproduced leading to the preliminary parametrization of the surface model for the fluence range $3\times10^{14}$ - $1.4\times10^{15}$ n$_{\textrm{\small eq}}$cm$^{-2}$. 

For more complete modeling of the CCE($x$) charge collection data of varying pitches of both strip and pixel detectors as well as higher number of fluence points will be required. Future efforts of the defect model developments within RD50 Collaboration also includes further calibration with the measured edge-TCT data that enables the tuning of the simulated $E$($z$) profile. The ultimate goal of the CCE($\Phi$) simulations is to stretch the defect models up to $\sim$$2\times10^{16}$ n$_{\textrm{\small eq}}$cm$^{-2}$ to account for the HL-LHC fluences of the pixel and 3D columnar detectors. 